\documentclass[aps,pra,longbibliography,nofootinbib,amsthm,amsmath,amssymb,amsfonts,notitlepage]{revtex4-1}
\usepackage[utf8]{inputenc}
\usepackage[T1]{fontenc}
\usepackage[english]{babel}
\usepackage{graphicx}
\usepackage{float}
\usepackage{cleveref}
\usepackage{xcolor}
\usepackage{tikz}

\newcommand{\bk}{{\bf k}}
\newcommand{\bb}{{\bf b}}
\newcommand{\bP}{{\bf P}}
\newcommand{\bp}{{\bf p}}
\newcommand{\br}{{\bf r}}
\newcommand{\bbe}{{\bf e}}

\def\lsim{\mathrel{\rlap{\lower4pt\hbox{\hskip1pt$\sim$}}
		\raise1pt\hbox{$<$}}}         

\def\gsim{\mathrel{\rlap{\lower4pt\hbox{\hskip1pt$\sim$}}
		\raise1pt\hbox{$>$}}}         

\date{\today}

\begin{document}

\title{Observability of the superkick effect within a quantum-field-theoretical approach}
\author{Igor P. Ivanov}
\email{ivanov@mail.sysu.edu.cn}
\author{Bei Liu}
\email{liub98@mail2.sysu.edu.cn}
\author{Pengming Zhang}
\email{zhangpm5@mail.sysu.edu.cn}
\affiliation{School of Physics and Astronomy, Sun Yat-sen University, 519082 Zhuhai, China}

\begin{abstract}
An atom placed in an optical vortex close to the axis may, upon absorbing a photon, acquire a transverse momentum 
much larger than the transverse momentum of any plane-wave component of the vortex lightfield. 
This surprising phenomenon dubbed superkick has been clarified previously in terms of the atom wave packet evolution
in the field of an optical vortex treated classically.
Here, we study this effect within the quantum field theoretical (QFT) framework.
We consider collision of a Bessel twisted wave with a compact Gaussian beam focused to a small focal spot $\sigma$
located at distance $b$ from the twisted beam axis.
Through a qualitative discussion supported by exact analytical and numerical calculations,
we recover the superkick phenomenon for $\sigma \ll b$
and explore its limits when $\sigma$ becomes comparable to $b$.
	On the way to the final result within the QFT treatment, we encountered and resolved
apparent paradoxes related to subtle issues of the formalism.
These results open a way to a detailed QFT exploration of other superkick-related effects 
recently suggested to exist in high-energy collisions.
\end{abstract}

\maketitle

\section{Introduction}

Physics of structured light---that is, a nonplane-wave light field with phase singularities
and other topological features---is a fascinating and actively studied topic 
of modern optics \cite{rubinsztein2016roadmap,babiker2019atoms,forbes2021structured}.
A prototypical example of structured light is an optical vortex: 
the electromagnetic field configuration near a singularity line
whose phase depends on the azimuthal variable $\varphi$ as $\exp(im \varphi)$, with an integer winding number $m$.
Such a light field is readily available in the focal region of the Laguerre-Gaussian modes of a laser beam
known since early 1990's \cite{Allen:1992zz} and nowadays routinely used in numerous directions
of fundamental research and in applications \cite{andrews2012angular,Paggett:2017,Knyazev:2019}.

Optical vortex is characterized with a swirling Poynting vector density, which rotates around the phase singularity axis.
As a result, light with optical vortex carries nonzero orbital angular momentum (OAM) proportional to $m$.
If an atom is placed in the light field of an optical vortex, also called ``twisted light'', 
it experiences light-induced torque proportional to the optical OAM \cite{babiker1994}. 
Also, selection rules governing its internal transitions
are modified, and these modifications depend on its distance to the axis \cite{babiker2002orbital,Afanasev:2014,schmiegelow2016transfer,afanasev2018experimental,solyanik2019excitation,schulz2019modification}.
This observation is intuitively clear: since each photon of the optical vortex carries a nonzero OAM, 
it can enhance multipole transitions.

What is more surprising---and even counter-intuitive---is the prediction of \cite{barnett2013superkick}
that an atom placed in the vicinity of the optical vortex axis 
can acquire a much larger transverse momentum than any of the photons of the twisted light beam can deliver.
This unexpectedly large momentum transfer was called in \cite{barnett2013superkick} a ``superkick''.

Semiclassical arguments behind the superkick phenomenon were outlined in \cite{barnett2013superkick}
and investigated with specific examples in \cite{Afanasev:2020nur}.
Indeed, if we follow a circular path with radius $b$ around the phase singularity axis, we accumulate a phase change of $2\pi m$ 
along the circumference $2\pi b$. The resulting phase gradient can be interpreted as the local momentum \cite{Berry2013fivemomenta}.
If we choose unit vectors $\bbe_x$ and $\bbe_y$ on the transverse plane and consider the
point $\bb = b \bbe_x$, then the local phase gradient on this plane produces the local momentum
\begin{equation}
\bp_\perp = \frac{m}{b}\, \bbe_y\label{mb}\,,
\end{equation}
which is orthogonal to $\bb$. If $b$ is sufficiently small, this local momentum can be arbitrarily large and can be transferred to an atom 
placed sufficiently close to the axis. However, if we interpret the light-atom interaction in terms of photons,
each photon can carry much smaller transverse momentum. We arrive at a paradox: photon absorption 
induces much larger momentum transfer than it actually carries.

The semiclassical arguments outlined above blend in uncontrollable way quantum and classical concepts.
They rely on pointlike classical atoms, which are, nevertheless, supposed to absorb photons, quantum entities. 
In \cite{barnett2013superkick}, Barnett and Berry made one step forward and represented the atom as a compact wave packet
of spatial extent $\sigma$. The smallness of $\sigma$ implies that, in momentum space, the wavefunction 
of this wave packet extends up to the large values of $\approx 1/\sigma$.
In the initial wave packet, all these plane-wave components balance each other leading to $\langle\bk_\perp\rangle = 0$.
However, when distorted by the interaction, this set of plane-wave components can easily produce a nonzero, large
average transverse momentum.
This picture offers the qualitative resolution to the paradox:
the photon does not supply large transverse momentum to the atom, but only rearranges and reweights the plane-wave components
inside the wave packets which produce the superkick.
In \cite{Berry2013fivemomenta}, Berry also interpreted this effect in terms of ``superweak'' values
which arise in experiments with postselection and can exceed the limits of the true spectrum of the operator.

This explanation, although enlightening, is still only partially quantum, as the atom wave packet 
is supposed to evolve in the classical external lightfield of the optical vortex.
Although it may be sufficient for this particular problem, one may want to explore similar effects
in high-energy physics collisions, and for that task, a full quantum field theoretic approach is needed.

The suggestion that superkick related effects may indeed occur and be observable in nuclear and high-energy physics realm 
was put forth in the very recently papers \cite{Afanasev:2020nur,Afanasev:2021fda}.
A significant shift of the energy threshold as well as other phenomena were predicted in vicinity of the twisted photon axis. 
However the analysis in these publications followed the formalism used for light-atom
interaction in optical vortices.

The purpose of the present paper is to address the superkick effect---and other puzzling observations 
which arise on the way---in the quantum field theoretic (QFT) framework.
We will reformulate the process as a QFT scattering problem, in which 
a twisted (Bessel) photon collides with a compact Gaussian wave packet 
representing the counterpropagating particle offset by an impact parameter $b$.
We will compute the expectation value of the transverse momentum of the final state
and explore its dependence on the parameters of the collision process.
Through exact analytical results, qualitative discussions, and numerical calculations, 
we will clarify all the features of this process which seem paradoxical and, eventually, 
place limits on the superkick observability in typical high-energy collisions.

	Recovering the superkick phenomenon and resolving the subtle technical issues which arise when
treating this problem within the QFT formalism with monochromatic beams provide, eventually,
useful insights of how to treat such problems within the full QFT framework.
Thus, we view this work as a first step towards a detailed systematic study,
and the experience gained here will be instrumental in adapting the formalism 
and exploring additional effects.

The structure of this paper is the following. In the next Section, we will describe
in more details the semiclassical roots of the superkick phenomenon.
We will also describe how we are going to treat this problem in the QFT framework,
highlighting yet another seemingly paradoxical observation.
Then in Section~\ref{section-nonplane-wave} we will outline the necessary QFT formalism
and resolve the second paradox just mentioned.
The following Section~\ref{section-Bessel-Gauss} analyzes the kinematics of 
the Bessel-Gaussian collision.
These results will finally lead us in Section~\ref{section-superkick} 
to the superkick phenomenon and its limits.
With these results, we give our comments on previous works in Section~\ref{section-comments}
and finally draw conclusions.
Appendices contain further technical calculations used in the main text.
Vectors are denoted with bold symbols and relativistic units $\hbar = c = 1$ are used throughout the paper. 

\section{The semiclassical origin of the superkick}\label{section-semiclassical}

\begin{figure}[!h]
	\centering
	\includegraphics[width=0.3\textwidth]{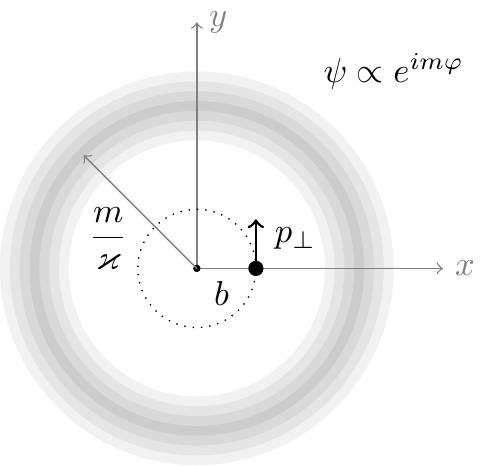}
	\caption{An atom placed at distance $b \ll m/\varkappa$ from the Bessel photon axis can acquire the transverse momentum $p_\perp \gg \varkappa$.}
\label{fig-semiclassical}
\end{figure}

Consider a Bessel photon, that is, a cylindrical monochromatic solution to Maxwell's equations
with a definite energy $\omega$, a definite longitudinal momentum $p_z$, 
and a definite modulus of the transverse momentum $|\bp_\perp|=\varkappa$.
Within the paraxial approximation, one can treat spin and OAM separately \cite{Allen:1992zz}.
Since we focus on the kinematic features of the process, from now on we will suppress the polarization vector
of the photon and discuss only its coordinate or momentum-space wavefunction.
The Bessel photon in the coordinate space can be represented as
\begin{equation}
\psi_{p_z, \varkappa, m}(\rho, z) \propto e^{ip_z z} e^{im\varphi} J_{|m|}(\varkappa \rho)\,.
\end{equation}
Here, $\rho = |\br_\perp|$, with the point $\rho = 0$ corresponding to the phase singularity.
The transverse intensity profile has the shape of concentric rings, 
with the first ring having radius $\rho = m/\varkappa$, see Fig.~\ref{fig-semiclassical}.

Consider now a circle of radius $b$ circumscribed in the transverse plane around the point of phase singularity.
The accumulated phase change of $2\pi m$ along the circumference $2\pi b$ produces
the phase gradient $m/b$, which leads to the local transverse momentum value \eqref{mb}.
At the first intensity ring, $b = m/\varkappa$, and we get $p_\perp(b)=\varkappa$,
which coincides with the transverse momentum carried by every plane wave component of the Bessel photon.
However, for smaller $b$, one gets the local transverse momentum arbitrarily larger than $\varkappa$.
Of course, at $b \ll m/\varkappa$, the intensity of the light field is suppressed, 
so that the probability of absorbing a photon is very small.
But when the atom does absorb the photon, it is predicted to acquire a superkick, 
a recoil transverse momentum much larger than $\varkappa$.
Thus, we confirm the apparent paradox: a superposition of plane waves with fixed
$|\bp_\perp|=\varkappa$ is able to deliver a transverse momentum (much) larger than $\varkappa$.

One may argue that the local momentum density defined via the local phase gradient 
is not the same thing as the momentum of the twisted photon.
Since the twisted state is not a plane wave, it is not a transverse momentum eigenstate.
One could then think of the expectation value of the transverse momentum operator calculated for the Bessel photon state.
However, due to the azimuthal symmetry of the Bessel twisted state, this expectation value is zero: 
$\langle \bp_\perp \rangle = 0$.

Thus, trying to make sense of the process, we encounter another puzzling observation:
an initial state (the photon and the atom) with a zero average transverse momentum
goes into a final state (an excited atom or a scattered system) 
with a nonzero---and potentially large---transverse momentum.
How does it agree with momentum conservation which must hold at the fundamental level?

Notice that this question would not arise if we considered a wave packet evolving in
an external field, as in \cite{barnett2013superkick}. It is only in the empty space collision setting
that we can appeal to the momentum conservation law and present its nonconservation as a paradox.

To clarify all these observations and resolve the paradoxes, we will consider the process in the 
full QFT setting. We will analyze free-space scattering of a monochromatic Bessel twisted state 
with parameters $\varkappa$ and $m$
with a monochromatic Gaussian beam of transverse spatial extent $\sigma$ localized at the impact parameter $b$ from the axis.
Replacing a semiclassical pointlike particle with a wave packet of size $\sigma$ will allows us to
check how the average transverse momentum of the final system changes as we vary $\sigma$ and $b$ with respect to $m/\varkappa$.
Also, by varying the ratio $\sigma/b$, we will be able to see how the superkick phenomenon
which exists for $\sigma \ll b$ gradually disappears for $\sigma \sim b$.

\section{Scattering of nonplane-wave states} \label{section-nonplane-wave}
\subsection{General considerations}

	Computation of non-planewave states scattering 
	within the QFT framework does not represent, by itself, a conceptual novelty.
	In fact, many QFT textbooks, when deriving the planewave scattering amplitude, first regularize the
	intermediate expressions with localized wavefunctions for the initial and final states 
	and then take the limit of plane waves, see e.g. \cite{Peskin}. Since the scattering amplitude is a linear functional
	of these initial wavefunctions, the construction of QFT itself is not altered, so that one can re-use the planewave 
	scattering matrix amplitude by inserting it between the appropriate wavefunctions.
	
	However, the kinematical distributions change significantly, especially when dealing with vortex states of non-zero OAM $m$.
	Indeed, some of the energy and momentum delta functions are integrated out when performing convolution 
	with the initial wavefunctions and do not appear anymore in the final particles phase space. 
	As a result, one observes novel distributions in the final state kinematics, which are unattainable for planewave collisions.
	The superkick phenomenon studied here also belongs to the family of kinematical peculiarities driven by the non-planewave nature 
	of the colliding states. This is why we focus in this work on universal kinematical features, 
	not on specific particles which collide.
	
	It should be kept in mind that the way these kinematical features are derived 
	differs significantly within a semiclassical treatment, in quantum mechanical scattering on a fixed scattering center,
	which can absorb any momentum transfer, 
	and in the QFT description where all initial states are represented as wavepackets 
	and the total energy-momentum conservation law is imposed.
	Therefore, even if a phenomenon is demonstrated within a semiclassical approach,
	it is desirable to rederive it within the full QFT.
	This derivation will shed more light on possible limitations of the effect 
	and on possible subtle issues. 
	Also, it can be a step towards a systematic study of other particle production processes
	which require the QFT framework.

With these reservations in mind, we begin with a brief reminder of the general QFT approach 
to calculating collision of free propagating quantum particles.
To illustrate the main kinematic effects, we consider elastic scattering of spinless particles.
The inclusion of the polarization degrees of freedom for fermions or vector bosons may lead to interesting additional effects
but they are not essential in our discussion of the superkick phenomenon and the apparent kinematic paradoxes outlined above.
 
First, we consider the textbook case, in which two plane wave states 
with three-momenta $\bk_1$, $\bk_2$ and energies $E_1$, $E_2$  
scatter to two final-state particles with three-momenta $\bk'_1$, $\bk'_2$ and energies $E'_1$, $E'_2$.
With a slight abuse of notation, we denote the total initial momentum as $\bP_0 = \bk_1 + \bk_2$, 
the total energy as $E_0 = E_1+E_2$, and the total final momentum as $\bP = \bk'_1 + \bk'_2$.
The plane wave $S$-matrix element has the form
\begin{equation}
S_{PW}(k_1,k_2;k'_1,k'_2) = i(2\pi)^4\delta(E_0-E'_1-E'_2) \delta^{(3)}(\bP_0 - \bP) {{\cal M} \over \sqrt{16 E_1 E_2 E'_1 E'_2}}\,.
\label{SPW}
\end{equation}
Here, ${\cal M}$ is the plane-wave invariant amplitude calculated according to the standard Feynman rules.
It can depend on kinematic invariants involved; 
in the simplest case of the pointlike quartic interaction, it is just a constant.
Squaring this amplitude, regularizing the squares of delta-functions in the usual way, 
one gets the differential cross section 
\begin{equation}
d\sigma_{PW} \propto  \delta(E_0-E'_1-E'_2) \delta^{(3)}(\bP_0 - \bP) |{\cal M}|^2  \, 
\frac{d^3 \bk'_1}{2E'_1} \frac{d^3 \bk'_2}{2E'_2}\,.\label{sigma-PW}
\end{equation}
Clearly, for fixed initial $\bk_1$ and $\bk_2$, the total final state momentum $\bP$ is also fixed.
The plane wave cross section can only display a non-trivial distribution in $\bk_1'$ or $\bk_2'$ but not in $\bP$.

Let us now assume that the two initial particles are prepared in nonplane-wave states;
for the final particles we still use the plane wave basis.
Scattering theory of arbitrarily shaped, partially coherent beams
was developed in the paraxial approximation in \cite{Kotkin:1992bj}.
The most general formalism, capable of going beyond the paraxial approximation, 
was presented in \cite{Karlovets:2016jrd,Karlovets:2020odl}.
The formalism was applied to twisted particle scattering in 
\cite{Jentschura:2010ap,Jentschura:2011ih,Ivanov:2011kk,Karlovets:2012eu,Ivanov:2016oue,Karlovets:2016jrd,Karlovets:2020odl}.
For the present analysis, it suffices to stick to the paraxial approximation and consider both initial particles as monochromatic states
with energies $E_1$ and $E_2$, as before, but with non-trivial momentum space wavefunctions 
$\phi_1(\bk_1)$ and $\phi_2(\bk_2)$.
The exact normalization condition for these wavefunctions is a subtle issue \cite{Jentschura:2011ih,Ivanov:2011kk,Ivanov:2011bv,Karlovets:2012eu}.
However, it is inessential to our discussion because we will be interested in the average values of operators, 
not the absolute magnitude of the cross section.
With these reservations, we can write the $S$ matrix element for scattering of the initial nonplane-wave state $i$
to the final plane wave state as 
\begin{equation}
S(i; k'_1,k'_2) = N \int d^3k_1 d^3k_2\, \phi_1(\bk_1) \phi_2(\bk_2) \cdot S_{PW}(k_1,k_2;k'_1,k'_2)\,,\label{SS}
\end{equation}
where the (inessential) factor $N$ takes care of all the normalization conditions.

We remark in passing that, up to normalization, this very quantity $S(i; k'_1,k'_2)$ can be viewed as the momentum space wavefunction 
of the final two-particle system $\Psi_f(\bk'_1,\bk'_2)$.
It is the same object as the evolved wavefunction discussed very recently in \cite{Karlovets:2021gcm}, that is, 
the wavefunction emerging from the scattering process itself and freely expanding in vacuum before hitting the detector.
In general, $\Psi_f(\bk'_1,\bk'_2)$ can depend on $\bk'_1$ and $\bk'_2$ individually, 
but in the simplest case of pointlike interaction it depends only on the total momentum $\bP$ and total energy $E_1' + E_2'$.

The plane-wave $S$-matrix element \eqref{SPW} contains the four-dimensional kinematic delta-function.
Since we deal with monochromatic initial states, the energy delta function remains unchanged 
and can be taken out of the integral \eqref{SS}.
Then, squaring this expression, regularizing the energy delta-function squared as before,
we can write the cross section in the following generic way:
\begin{equation}
d\sigma \propto |{\cal I}|^2  d\Gamma'\,, \quad \mbox{where}
\quad d\Gamma' = \delta(E_0-E'_1-E'_2) \, \frac{d^3 \bk'_1}{2E'_1} \frac{d^3 \bk'_2}{2E'_2}\,,\label{sigma-general}
\end{equation}
while the function ${\cal I}$ includes the integration over all plane wave components of the two initial states:
\begin{equation}
{\cal I} = \int d^3\bk_1 d^3\bk_2\, \phi_1(\bk_1) \phi_2(\bk_2)\, \delta^{(3)}(\bk_1+\bk_2-\bP)\cdot {\cal M}\,.\label{main-1}
\end{equation}
The proportionality symbol in \eqref{sigma-general} refers to the normalization coefficients whose accurate analysis is inessential.

Let us examine the key features of the general nonplane-wave cross section $d\sigma$ in \eqref{sigma-general}.
Unlike the plane wave case where the fixed initial momenta forced the total final $\bP$ to be fixed,
here we have a distribution in $\bP$. This is illustrated by changing the final phase space
from $d^3 \bk'_1 d^3 \bk'_2$ to $d^3 \bk'_1 d^3 \bP$ and observing the absence of $\delta^{(3)}(\bk_1+\bk_2-\bP)$ in \eqref{sigma-general}.
This distribution is a new dimension in the final state kinematics which was absent in the pure plane-wave scattering
and which can now be explored.

Further simplifications are possible in the realistic situation when $\phi_1(\bk_1)$ and $\phi_2(\bk_2)$
are compact wavefunctions localized around the average momenta $\bp_1 = \langle \bk_1 \rangle$ and $\bp_2 = \langle \bk_2 \rangle$.
We assume that these two average momenta are antiparallel to each other, thus defining the common axis $z$. 
We write them as $\bp_1 = (0,0,p_{1z})$, $\bp_2 = (0,0,p_{2z})$, with the convention that $p_{1z} > 0$, 
$p_{2z} = -|p_{2z}| < 0$. We denote their sum as $P_{z0} = p_{1z} + p_{2z}$.
Although the individual final particle momenta $\bk'_1$ and $\bk'_2$ can be large,
the total final transverse momentum $\bP_\perp$ as well as the deviation of the final longitudinal momentum from $P_{z0}$
\begin{equation}
\Delta P_z = P_z - P_{z0} = P_z - p_{1z} + |p_{2z}|\label{DeltaPz}
\end{equation}
are expected to be small.

We use the integration over $dk_{1z}$ to eliminate the energy delta function. 
As we show in Appendix~\ref{appendix-2particles}, for the compact initial momentum-space wavefunctions, 
the vast majority of the final phase space 
$d\Gamma'$ can be presented in the factorized form separating the total momentum and the relative motion:
\begin{equation}
d\Gamma' =  \sum_{k^*_{1z}} \frac{d^2 \bk'_{1\perp}}{2\sqrt{\lambda}} \cdot d^2 \bP_\perp \, d\Delta P_z\,.\label{dGam0}
\end{equation}
where the function $\lambda$ is depends on $\bk_{1\perp}^{\prime 2}$ and $P_{0z}$ but not on $\bP_\perp$ or $d\Delta P_z$.
The summation here refers to the two possible values of $k'_{1z}$ available for a given $\bk'_{1\perp}$ and $\bP$.

Using this approximate factorization of the final phase space and recalling 
that ${\cal I}$ also depends on $\bP$ but not on $\bk'_{1\perp}$, 
we can compute the average transverse momentum of the final state as
\begin{equation}
\langle \bP_\perp\rangle = \frac{\langle\Psi_f|\bP_\perp|\Psi_f\rangle}{\langle\Psi_f|\Psi_f\rangle} = 
\frac{\int |{\cal I}|^2 \bP_\perp d\Gamma'}{\int |{\cal I}|^2 d\Gamma'} \approx 
\frac{\int |{\cal I}|^2 \bP_\perp d^2 \bP_\perp \, d\Delta P_z}{\int |{\cal I}|^2 d^2 \bP_\perp \, d\Delta P_z}\,,
\label{average-P-def}
\end{equation}

\subsection{The momentum non-conservation paradox and its resolution}

Before we proceed with the computation of $\langle \bP_\perp\rangle$, 
let us address the apparent non-conservation of the transverse momentum mentioned in Section~\ref{section-semiclassical}.
Indeed, with the above convention for axis $z$, we see that the average transverse momentum of the initial state 
is zero $\bp_{1\perp}+\bp_{2\perp} = 0$.
However, we are computing the final transverse momentum \eqref{average-P-def} 
and expect it to be nonzero.
How can it happen, if momentum is conserved?

\begin{figure}[!h]
	\centering
	\includegraphics[width=0.4\textwidth]{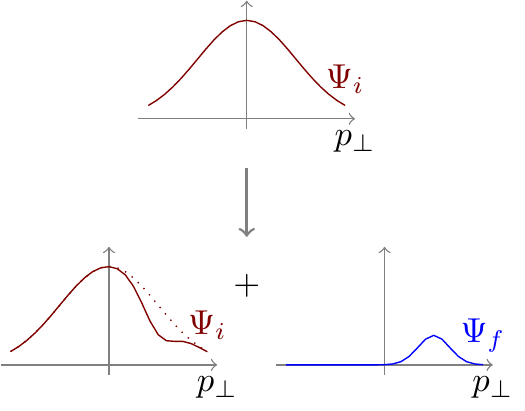}
	\caption{Schematic illustration of the origin of a nonzero transverse momentum in the scattered part of the final wavefunction.}
	\label{fig-pt-origin}
\end{figure}

The answer is that collision of two incoming waves leads not only to the scattered state we have considered 
but also to a nonscattered initial state which is only partially depleted by the interaction, see Fig.~\ref{fig-pt-origin}. 
The full outgoing wavefunction is a superposition of these components, which even belong to different Fock states
if the collision is inelastic.
The $S$-matrix element presented above describes only the scattered component $\Psi_f$ of the total final wavefunction.

Depending on the wavefunctions and on the fundamental interactions, it may happen
that the intensity of the scattering process exhibits certain bias in the transverse momentum space. 
As a result, the scattered component of the outgoing wavefunction may get a nonzero $\langle \bP_\perp\rangle$,
while the depleted nonscattered component acquires a nonzero compensating average momentum.
The transverse momentum of the {\em entire} final state, 
which includes both scattered and nonscattered components, is zero.
Thus, scattering does not produce transverse momentum out of nothing; it just rearranges the individual plane-wave components.

In the usual plane wave scattering, we never run into this paradox because 
the momentum space wavefunction of the initial state is assumed to be infinitely narrow: $\phi(\bk) \propto \delta^{(3)}(\bk-\bp)$.
There is simply no room to induce momentum bias within this infinitely narrow momentum distributions.
Such effects can arise only with non-trivially shaped initial states $\varphi(\bk_i)$.
Even an exact Bessel beams would suffice. Although in such a state the values of $k_z$ and $|\bk_\perp|$
are fixed, the azimuthal angle distribution of $\varphi(\bk_i)$, when disturbed, can produce a nonzero overall transverse momentum.

\section{Scattering of Bessel twisted state and Gaussian beam}\label{section-Bessel-Gauss}

\subsection{The setting}

Now we focus on scattering of a Bessel twisted state 
with a Gaussian beam whose center is shifted from the twisted state beam axis by $\bb_\perp$. 
The Bessel twisted state is defined by a fixed longitudinal momentum $k_{1z} = p_{1z} > 0$, 
a fixed modulus of the transverse momentum $|\bk_{1\perp}| = \varkappa$, and the OAM $m$,
see \cite{Jentschura:2010ap,Jentschura:2011ih,Ivanov:2011kk,Karlovets:2012eu} for more details.
The momentum space wavefunction $\phi_1(\bk_1)$, up to normalization, is then given by
\begin{equation}
\phi_1(\bk_1) = \delta(k_{1z} - p_{1z})\delta(|\bk_{1\perp}| - \varkappa) e^{i m \varphi_1}\,.\label{twisted1}
\end{equation}
To construct the second initial particle wave packet, we begin with the plane-wave state with momentum $\bp_2 = (0,0,p_{2z})$,
where $p_{2z} < 0$ satisfies the energy-momentum relation $E_2^2 = p_{2z}^2 + M^2$. 
We construct the Gaussian beam by smearing the momentum distribution
around $\bp_2$ and keeping the beam monochromatic:
\begin{equation}
\phi_2(\bk_2) = \exp\left[-\frac{(\bk_2 - \bp_2)^2\sigma^2}{2}\right] \delta(\bk_2^2 + M^2 - E_2^2)\cdot e^{-i\bk_{2\perp}\bb_\perp}\,.\label{gauss1}
\end{equation}
Here, the parameter $\sigma$ indicates the transverse spatial extent of the wavefunction in the coordinate space,
while the phase factor with $\bb_\perp$ describes a shift of the beam with respect to the twisted particle axis.
In the calculations below, we take the paraxial approximation for the Gaussian state: 
\begin{equation}
\sigma |p_{2z}| \gg 1\,.\label{paraxial}
\end{equation}

To make it clear, we do not claim that the expression \eqref{gauss1} is the best choice for description
of realistic wave packets. It is explicitly non-Lorentz-invariant, which by itself does not represent any problem.
Indeed, nonplane-wave states can be produced in an experimental setting which, by definition,
resides in a certain reference frame.
Second, one could certainly consider other monochromatic transversely localized beams 
as well as 3D localized wave packets which are unavoidably non-monochromatic.
One could also choose different longitudinal and transverse sizes $\sigma_z \not = \sigma_\perp$
or consider a non-Gaussian longitudinal profile. As long as we stay within the paraxial approximation,
these modifications will differ by minor details of the longitudinal momentum distribution of the cross section.
All the transverse space phenomena, in particular the phenomenon of superkick,
are expected to be insensitive to these details.

Thus, we deal with a four length scale problem: we have the wavelength of the Gaussian state $\lambda = 1/|p_{2z}|$,
its transverse wave packet size $\sigma$, its distance to the twisted photon axis $b$, and the radius of the first Bessel beam ring 
in the transverse plane $m/\varkappa$.
The kinematic configuration appropriate for observation of the superkick phenomenon implies the following hierarchy 
of these length scales:
\begin{equation}
\frac{1}{|p_{2z}|} \ll \sigma \ll b \ll \frac{m}{\varkappa}\,.\label{scales}
\end{equation}
When testing the limits of this configuration, we will also explore values $\sigma \sim b$ or even $\sigma \gg b$. 
Notice that, by taking $\sigma \to \infty$, we can recover the plane wave limit for the Gaussian state.
In this limit, the results should not depend on $b$, and the cross section should be represented as the 
azimuthal averaged value of the plane wave cross section, as was derived in \cite{Ivanov:2011kk}.

\subsection{Towards the superkick: an initial attempt}

With all these considerations, we substitute the initial state wavefunctions into the integral ${\cal I}$ of \eqref{main-1}: 
\begin{eqnarray}
{\cal I} &=& \int d^3\bk_1 d^3\bk_2 \, \delta(k_{1z} - p_{1z})\delta(|\bk_{1\perp}| - \varkappa) e^{i m \varphi_1}\cdot 
\, e^{-\frac{(\bk_2 - \bp_2)^2\sigma^2}{2}} \delta(\bk_2^2 + M^2 - E_2^2)\, e^{-i\bk_{2\perp}\bb_\perp} 
\delta^{(3)}(\bk_1 + \bk_2 - \bP) \cdot {\cal M}\,.\quad\label{main-2}
\end{eqnarray}
It contains six integrations and six delta function and, therefore, can be calculated exactly, 
see details in Appendix~\ref{appendix-absence}.
As we calculated in that Appendix, delta functions place constraints on the possible values of $|\bP_\perp|$ and $\Delta P_z$,
which carve a crescent-like region on the $(\Delta P_z, |\bP_\perp|)$ plane.
In the small-$\Delta P_z$ region, where the cross section is expected to be concentrated, the borders of this region
can be represented by the two inequalities
\begin{equation}
(|\bP_\perp|-\varkappa)^2 < 2 |p_{2z}|\Delta P_z < (|\bP_\perp|+\varkappa)^2\,.\label{bP-Pz-plane} 
\end{equation}

\begin{figure}[!h]
	\centering
	\includegraphics[width=0.3\textwidth]{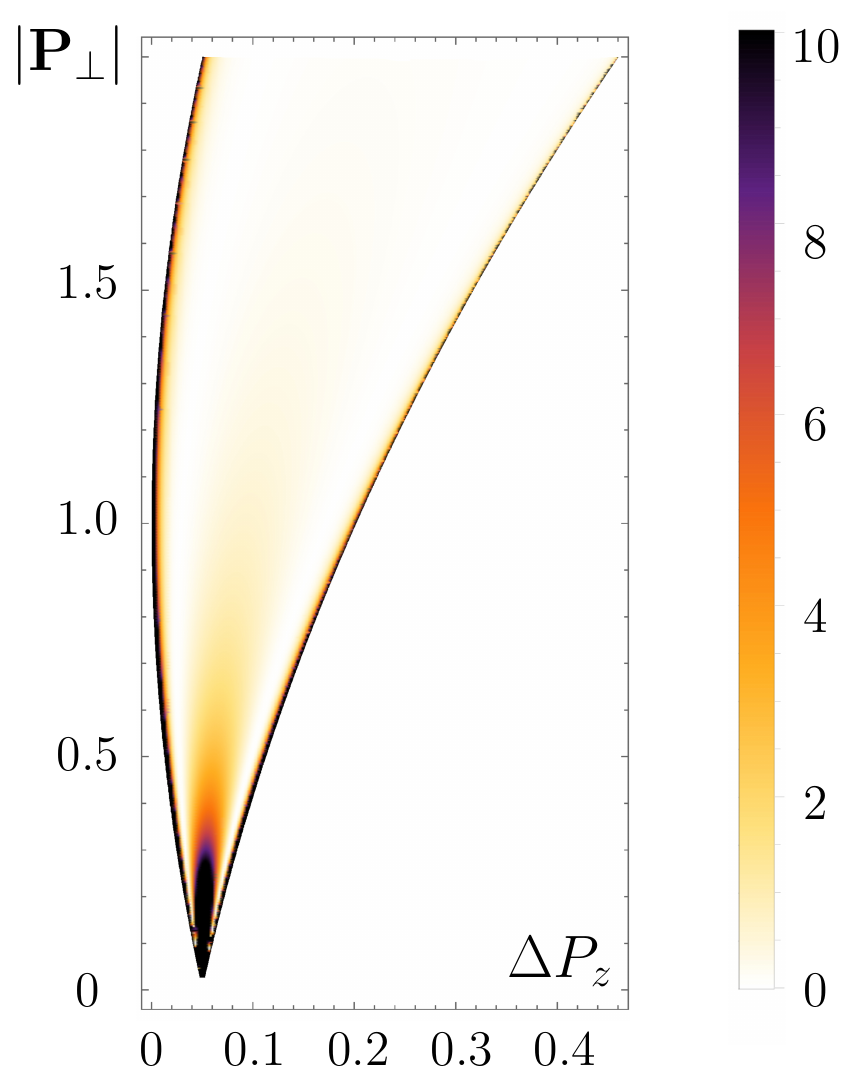}
	\caption{The $(|\bP_\perp|,\Delta P_z)$-distribution of the function $|{\cal I}|^2$, whose value is shown in arbitrary units and encoded in color shades. 
		The values of the parameters are (in units of an arbitrary momentum scale): 
		$|p_{2z}| = 10$, $\varkappa = 1$, $1/\sigma = 2$, $b = \sigma/5$, $m=2$.}
	\label{fig-crescent}
\end{figure}

The expression $|{\cal I}|^2$, which we compute exactly in Appendix~\ref{appendix-absence}, 
exhibits strong intensity oscillations inside this region and diverges near the boundaries. 
Apart from these small-scale ripples, it is concentrated at small values of $\bP_\perp$ and $\Delta P_z$
and exhibits exponential attenuation at their large values. 
The typical ranges of $\Delta P_z$ and $|\bP_\perp|$ are
\begin{equation}
\Delta P_z \lsim \frac{1}{\sigma^2 |p_{2z}|}\,, \quad
|\bP_\perp| \lsim \frac{1}{\sigma}\,,\quad
\Delta P_z \ll |\bP_\perp|\,.\label{DeltaPz-Pperp}
\end{equation} 
We conclude that the total momentum of the final state is predominantly transverse,
and it is driven by the transverse momenta of the initial compact Gaussian state, 
not by the transverse momentum of the Bessel twisted state.
	For the purpose of illustration, we show in Fig.~\ref{fig-crescent} the behavior of $|{\cal I}|^2$ 
	on the $(|\bP_\perp|,\Delta P_z)$-plane for $|p_{2z}| = 10$, $\varkappa = 1$, $1/\sigma = 2$ (all expressed
	in arbitrary but the same unit of momentum), $b = \sigma/5$, and $m=2$.

With these observations, we confirm the explanation of the origin of the superkick phenomenon given in \cite{barnett2013superkick}.
The initial compact wave packet {\em already} has large transverse momentum components of the order of $1/\sigma \gg \varkappa$.
However, these large transverse momentum components are neatly balanced in the initial Gaussian state, 
so that its averaged transverse momentum is zero.
The role of the twisted photon, which does not possess any significant transverse momentum, 
is just to {\em disturb this balance} in a particular, momentum-biased way. 
The outgoing scattered state still has large transverse momentum components
of the order of $1/\sigma$, but they can now lead to an overall nonzero average transverse momentum,
which can, of course, be greater than $\varkappa$.

Are these qualitative considerations confirmed by analytical results?
In Appendix~\ref{appendix-absence} we show that, if we proceed with the calculation of the 
average transverse momentum \eqref{average-P-def} using the exact expression for $|{\cal I}|^2$,
we run into a non-integrable divergence at the boundary.
This divergence is driven by the fact that the exact Bessel state is not normalizable
and requires a regularization scheme.
Following \cite{Jentschura:2010ap,Jentschura:2011ih,Ivanov:2011kk,Karlovets:2012eu},
one can normalize the Bessel state to a large but finite cylindrical volume
and render the integrals finite.
However in this case the average final transverse momentum momentum becomes parametrically suppressed
and, at small $b$, it is proportional to $b$ {\em itself}, not to $1/b$, 
as was expected from the pointlike probe semiclassical considerations.
Thus, this computation {\em does not} confirm the superkick phenomenon.

\begin{figure}[!h]
	\centering
	\includegraphics[width=0.5\textwidth]{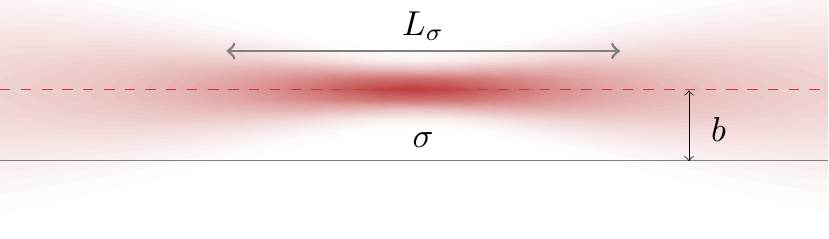}
	\caption{Transverse spreading of the Gaussian state along $z$ axis. The gray solid line corresponds to the axis of the Bessel state.}
	\label{fig-z-range}
\end{figure}

The origin of this new discrepancy becomes evident if we consider the scattering process in coordinate space.
The transverse distribution of the Bessel twisted state is, by construction, $z$-invariant.
In contrast, the transverse distribution of the Gaussian state \eqref{gauss1} evolves along the $z$ axis:
\begin{equation}
|\psi(\br_\perp, z)|^2 \propto \exp\left[- \frac{\br_\perp^2}{\sigma^2(z)}\right]\,, \quad
\sigma^2(z) = \sigma^2\left(1+\frac{z^2}{L_\sigma^2}\right)\,, \quad
\mbox{where}\quad L_\sigma = \sigma^2|p_{2z}|\,.\label{diff}
\end{equation}
The Gaussian beam stays well focused only over the longitudinal focal length $|z| \lsim L_\sigma$
and broadens at larger values of $|z|$, Fig.~\ref{fig-z-range}. It is clear that the superkick phenomenon is expected only
when $\sigma(z)$ stays much smaller than $b$, that is, over a limited $z$ interval.
Beyond this $z$ interval, the Gaussian state is unable to locally probe 
the very strong phase gradient of the twisted state.
When computing $\langle \bP_\perp\rangle$ of the final state,
we effectively performed the integration over all values of $z$ up to the size of the normalization volume. 
The Gaussian state was spreading in the transverse plane at large $|z|$, 
but its interaction with the Bessel twisted state never dies out 
because the Bessel state is not localized in the transverse plane.
Therefore, the resulting integral is dominated by very large $z$ values where no superkick can be expected.

We conclude that the absence of superkick which is detailed in Appendix~\ref{appendix-absence} 
is an artifact of the inappropriate setting we used so far. That is, 
by considering an exact monochromatic Bessel vs. Gaussian beam collision
in the {\em entire} coordinate space, we inadvertently dilute the effect
which is expected to exist only within a short $z$ range.

\subsection{Towards the superkick: limiting the $z$-range}

In order to remove the artifact and to recover the superkick phenomenon within the QFT framework,
we need to modify the collision setting to make sure that scattering takes place only 
where the Gaussian state is well focused.
Several approaches can be used. One is to pass from monochromatic beams to non-monochromatic, 
longitudinally localized wave packets.
This method is especially appropriate for interaction of atoms just released from a trap 
with a short twisted laser pulse.
Another approach is to replace the Bessel twisted state 
with Laguerre-Gaussian beams which are normalized in the transverse plane \cite{Karlovets:2020odl}.
In this way, the twisted state itself will be subject to defocusing, and the interaction will be 
saturated by the overlapping focal spots of the two beams.

These schemes will make the calculations cumbersome and, perhaps, hinder the essence of the effect.
To keep computations clear, we adopt here yet another prescription.
We keep the same exact monochromatic Bessel and Gaussian beams but introduce an auxiliary regulating
function $f(z)$ which switches on the scattering within the focal length, with $f(z) = 1$ at $z=0$,
and switches off for $|z|\gsim \ell$, well before the Gaussian state begins to significantly defocus.
For concreteness, we take the following regulating function:
\begin{equation}
f(z) = \exp(-|z|/\ell)\quad \mbox{with}\quad \sigma \ll \ell \ll L_\sigma\,.\label{ell-choice}
\end{equation}
The motivation for the inequality $\sigma \ll \ell $ will become clear below, while
the last condition guarantees that the integration is limited to the focal length of the beam, 
that is, the $z$-range where the Gaussian beam stays focused to its waist size $\sigma(z) \approx \sigma$.
The final result---that is, the value of the superkick and its dependence 
on the transverse parameters---should not depend on $\ell$.

The regulating function modifies the longitudinal momentum conservation in \eqref{main-2}.
Instead of 
$$
\delta(k_{1z}+k_{2z}-P_z) = \frac{1}{2\pi}\int dz\, e^{-i(k_{1z}+k_{2z}-P_z)z}
$$
we now have $g(k_{1z}+k_{2z}-P_z)$ defined as
\begin{equation}
g(k_{1z}+k_{2z}-P_z) = \frac{1}{2\pi}\int dz\, e^{-i(k_{1z}+k_{2z}-P_z)z} e^{-|z|/\ell}
= \frac{1}{\pi}\frac{\ell}{1+ (k_{1z}+k_{2z}-P_z)^2 \ell^2}\,.\label{gpz}
\end{equation}
As a result, the expression for ${\cal I}$ has one less delta function than in \eqref{main-2}.
The integration over $k_{2z}$ is now compensated not by $\delta(k_{1z}+k_{2z}-P_z)$
but by $\delta(\bk_2^2 - p_{2z}^2)$, which leaves the following $\Delta P_z$-depending functions:
\begin{equation}
{\cal I} \propto g(\tilde P_z)\, e^{-\tilde P_z^2 \sigma^2/2} \,, \quad \tilde P_z = \Delta P_z + \sqrt{p_{2z}^2-\bk_{2\perp}^2} - |p_{2z}|
\approx \Delta P_z - \frac{\bk_{2\perp}^2}{2|p_{2z}|} \,.
\end{equation}
The two functions---the regulating and the exponential functions---have different ranges.
The regulating function $g(\tilde P_z)$ stays flat for $|\tilde P_z| \lsim 1/\ell$ and then decreases.
Within this range, the argument of the exponential function stays below $\sigma^2/\ell^2 \ll 1$, where we used the first 
relation of \eqref{ell-choice}. Therefore, the exponential function can be safely replaced by 1 in front of $g(\tilde P_z)$.

\begin{figure}[!h]
	\centering
	\includegraphics[width=0.5\textwidth]{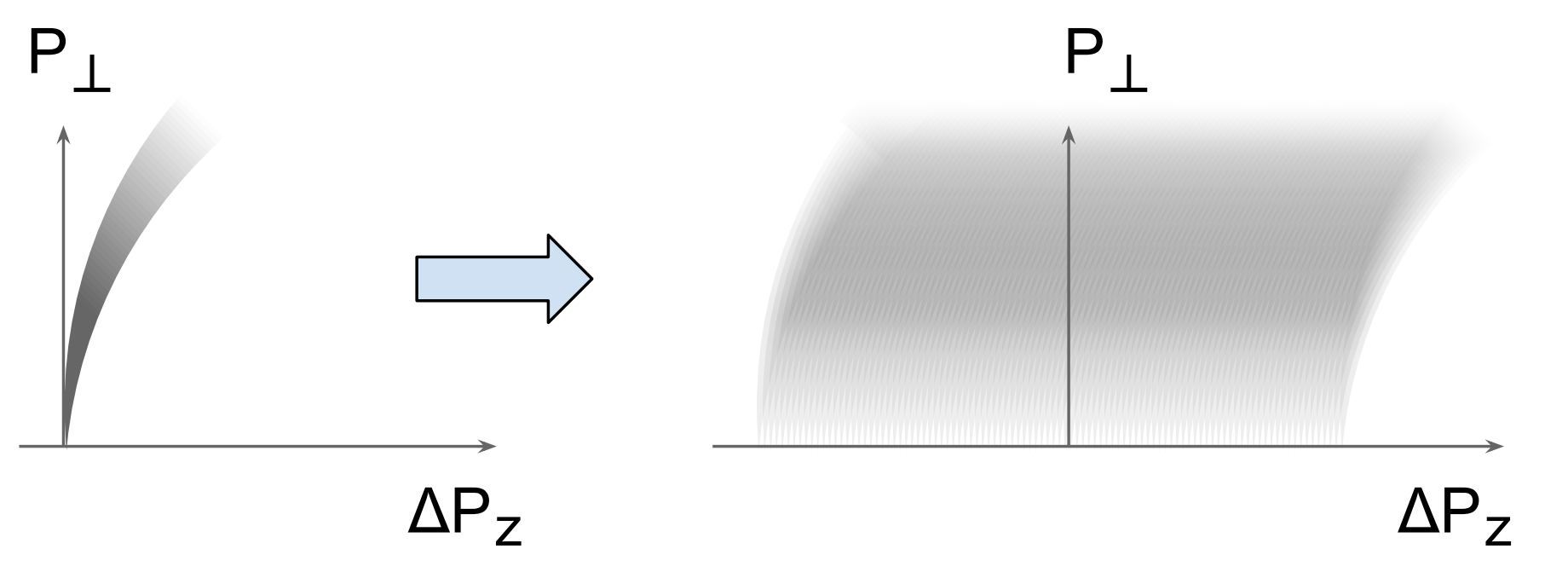}
	\caption{Schematical distribution of $|{\cal I}|^2$ on the $(|\bP_\perp|,\Delta P_z)$ plane upon restricting the $z$ integration to a distance
		within the focal length. Small scale oscillations and feature seen in Fig.~\ref{fig-crescent} are smoothed out upon smearing.}
	\label{fig-PtPz-smeared}
\end{figure}

The resulting $\Delta P_z$ distribution becomes much wider than before.
In \eqref{DeltaPz-Pperp}, we had $\Delta P_z \lsim 1/(\sigma^2|p_{2z}|) = 1/L_\sigma$.
Now, the $\Delta P_z$ distribution is governed by $g(\tilde P_z)$ and extends up to 
$1/\ell \gg 1/L_\sigma$, where we used the second inequality of \eqref{ell-choice}.
This effectively leads to a dramatic smearing of the entire picture of Fig.~\ref{fig-crescent} which is illustrated in Fig.~\ref{fig-PtPz-smeared}.

The end result is that ${\cal I}$ displays an approximately factorized dependence
on $\Delta P_z$ and $\bP_\perp$. The $\Delta P_z$ distribution comes mostly 
from $g(\tilde P_z) \approx g(\Delta P_z)$.
Therefore, when calculating the average final momentum \eqref{average-P-def}, 
one can factor both the numerator and the denominator into the transverse and longitudinal integrals 
and cancel the longitudinal ones.
Thus, the expression for the average final transverse momentum can be simplified to 
\begin{equation}
\langle \bP_\perp\rangle \approx 
\frac{\int |{\cal J}|^2 \bP_\perp d^2 \bP_\perp}{\int |{\cal J}|^2\, d^2 \bP_\perp}\,,
\label{average-P-2}
\end{equation}
where 
\begin{equation}
{\cal J} = \frac{1}{2\pi}\int d\varphi_1 \exp\left[i m \varphi_1 - i (\bP_\perp-\bk_{1\perp})\bb_\perp 
- \frac{(\bP_\perp-\bk_{1\perp})^2\sigma^2}{2}\right]\,.\label{kick-3}
\end{equation}
This is the expression that will display the superkick phenomenon in the appropriate 
kinematic setting and also will demonstrate its limits. We explore it in detail in the next Section.

Notice that, when defining ${\cal J}$, we assumed that the invariant amplitude ${\cal M}$
depends on $\bP_\perp$ very weakly and, therefore, can be factored out and canceled in \eqref{average-P-2}.
This is a very natural assumption for all non-resonant processes because the range of $\bP_\perp$ is very small 
compared to the individual final particles' momenta $\bk'_1$ and $\bk'_2$.
This assumption may be violated in the sharply resonant processes such as resonant photon scattering on atoms or ions.
Analysis of these subtleties is delegated to a future work. 

\section{Superkick and its limits}\label{section-superkick}

\subsection{Recovering the superkick}

Using integrals 3.937 from \cite{integrals}, we can formally express the integral ${\cal J}$ in \eqref{kick-3} as 
\begin{equation}
{\cal J} = e^{im\varphi_P}e^{-i\bP_\perp\bb_\perp - \frac{(P_\perp^2 + \varkappa^2)\sigma^2}{2}}
\left(\frac{P_\perp^2\sigma^4 + b^2e^{2i(\varphi_{b}-\varphi_P)}}{b^2+P_\perp^2\sigma^4-2bP_\perp \sigma^2 \sin(\varphi_{b}-\varphi_P)}\right)^{m/2}
I_{m}\left(\varkappa\sqrt{P_\perp^2\sigma^4 - b^2 - 2ibP_\perp\sigma^2 \cos\varphi_{bP}}\right)\,.\label{formal}
\end{equation}
Here, $\varphi_P$ and $\varphi_b$ are the azimuthal angles of $\bP_\perp$ and $\bb$, respectively.
Notice that in the limit $b \to 0$ all functions apart from the very first phase factor $e^{im\varphi_P}$ 
become real. This illustrates the OAM conservation for head-on, zero impact parameter collision.
However, this expression itself is not very enlightening, 
so instead of inserting it in the exact expression \eqref{average-P-2}, 
we find it instructive to perform the $\bP_\perp$ integration first in the numerator and denominator of this expression. 
The denominator becomes
\begin{eqnarray}
&&\int |{\cal J}|^2 d^2\bP_\perp =\int d^2 \bP_\perp \frac{1}{(2\pi)^2}\int d\varphi_1 d\varphi'_1
e^{im(\varphi_1  - \varphi'_1) + i (\bk_{1\perp} - \bk'_{1\perp})\bb_\perp}
\exp\left\{-\left[(\bP_\perp-\bk_{1\perp})^2 + (\bP_\perp-\bk'_{1\perp})^2\right]\frac{\sigma^2}{2}\right\}\nonumber\\
&&=\frac{1}{(2\pi)^2}\int d\varphi_1 d\varphi'_1 e^{im(\varphi_1  - \varphi'_1) + i (\bk_{1\perp} - \bk'_{1\perp})\bb_\perp}
\int d^2 \bP_\perp\exp\left[-\left(\bP_\perp - \frac{\bk_{1\perp} + \bk'_{1\perp}}{2}\right)^2\sigma^2
-\frac{(\bk_{1\perp}-\bk'_{1\perp})^2\sigma^2}{4}\right]\nonumber\\
&&=\frac{\pi}{\sigma^2}\cdot \frac{1}{(2\pi)^2}\int d\varphi_1 d\varphi'_1 
e^{im(\varphi_1  - \varphi'_1) + i (\bk_{1\perp} - \bk'_{1\perp})\bb_\perp}
\exp\left[-\frac{(\bk_{1\perp}-\bk'_{1\perp})^2\sigma^2}{4}\right]\,.\label{denom}
\end{eqnarray}
To recover the superkick regime, we consider $\sigma$ to be much smaller than $b$ and $1/\varkappa$. 
Then, the last factor here can be omitted as it always stays very close to 1.
As a result, the integrations with respect to $\varphi_1$ and $\varphi'_1$ decouple:
\begin{equation}
\int |{\cal J}|^2 d^2\bP_\perp =\frac{\pi}{\sigma^2}\cdot 
\frac{1}{2\pi}\int d\varphi_1 e^{im\varphi_1 + i \bk_{1\perp} \bb_\perp}\cdot
\frac{1}{2\pi}\int d\varphi'_1 e^{-im\varphi'_1 - i \bk'_{1\perp} \bb_\perp}
= \frac{\pi}{\sigma^2}\cdot \left[J_m(\varkappa b)\right]^2\,.\label{denom-1}
\end{equation}
For the numerator of \eqref{average-P-2}, we have an additional $\bP_\perp$ inserted.
Performing the $\bP_\perp$ integral as before, we get
$$
\int d^2 \bP_\perp\, \bP_\perp \,\exp\left[-\left(\bP_\perp - \frac{\bk_{1\perp} + \bk'_{1\perp}}{2}\right)^2\sigma^2\right]
= \frac{\pi}{\sigma^2}\cdot \frac{\bk_{1\perp} + \bk'_{1\perp}}{2}\,.
$$
Instead of Eq.~\eqref{denom-1}, we now have 
\begin{equation}
\int |{\cal J}|^2 \bP_\perp d^2\bP_\perp =\frac{\pi}{\sigma^2}\cdot 
\int d\varphi_1 d\varphi'_1 \frac{\bk_{1\perp} + \bk'_{1\perp}}{2} 
e^{im(\varphi_1  - \varphi'_1) + i (\bk_{1\perp} - \bk'_{1\perp})\bb_\perp}\,.\label{num-1}
\end{equation}
Let us define the unit vectors $\bbe_x$ and $\bbe_y$ along axes $x$ and $y$, with the axis $x$ chosen
along the impact parameter vector $\bb_\perp$ as in Fig.~\ref{fig-semiclassical}.
Then $\bk_{1\perp} = \cos(\varphi_1-\varphi_b)\bbe_x + \sin(\varphi_1-\varphi_b)\bbe_y$.
Substituting this expression into \eqref{num-1} and using the integrals
\begin{eqnarray}
\frac{1}{2\pi}\int d\varphi\, e^{im\varphi + i \varkappa b \cos\varphi}\cos\varphi &=& 
\frac{i^{m+1}}{2}[J_{m+1}(\varkappa b)-J_{m-1}(\varkappa b)] \,,
\nonumber\\
\frac{1}{2\pi}\int d\varphi\, e^{im\varphi + i \varkappa b \cos\varphi}\sin\varphi &=& 
\frac{i^{m}}{2}[J_{m+1}(\varkappa b)+J_{m-1}(\varkappa b)] = i^m \frac{m}{\varkappa b} J_m(\varkappa b)\,,
\end{eqnarray}
we obtain that only the $\bbe_y$ component survives:
\begin{equation}
\int |{\cal J}|^2 \bP_\perp d^2\bP_\perp =\frac{\pi}{\sigma^2}\cdot 
\frac{m}{b}[J_{m}(\varkappa b)]^2\, \bbe_y\,.\label{num-2}
\end{equation}
As a result, the transverse momentum of the final state, within the approximation $\sigma\ll b, 1/\varkappa$, 
is given by 
\begin{equation}
\langle \bP_\perp \rangle = \frac{m}{b} \, \bbe_y\,.\label{Pperp-3}
\end{equation}
This expression is valid for any value of $\varkappa b$.
Thus, we recovered the superkick phenomenon 
in the full quantum field theoretic computation of the Bessel twisted state collision 
with compact a Gaussian wave packet provided its size $\sigma\ll b, 1/\varkappa$.

\subsection{Exploring the limits of the superkick phenomenon}

\begin{figure}[!h]
	\centering
	\includegraphics[width=0.45\textwidth]{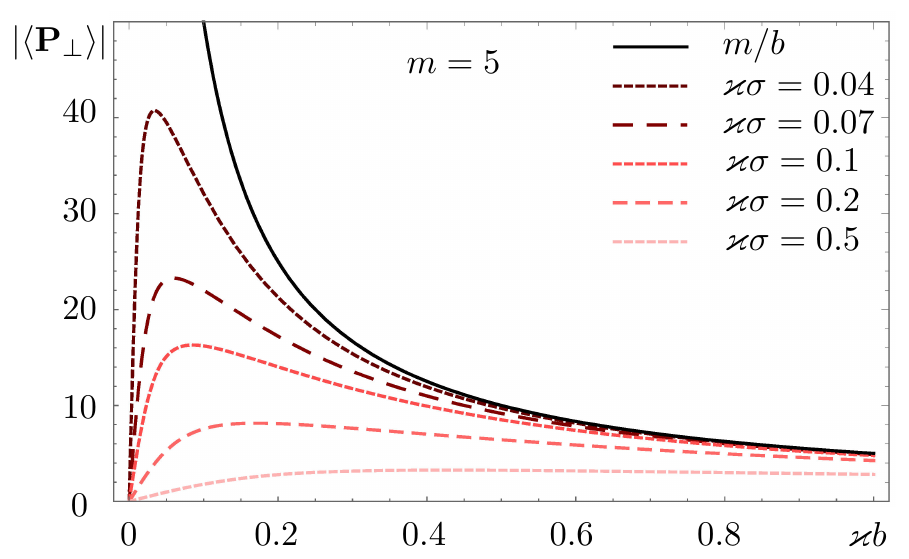}\\[2mm]
	\includegraphics[width=0.45\textwidth]{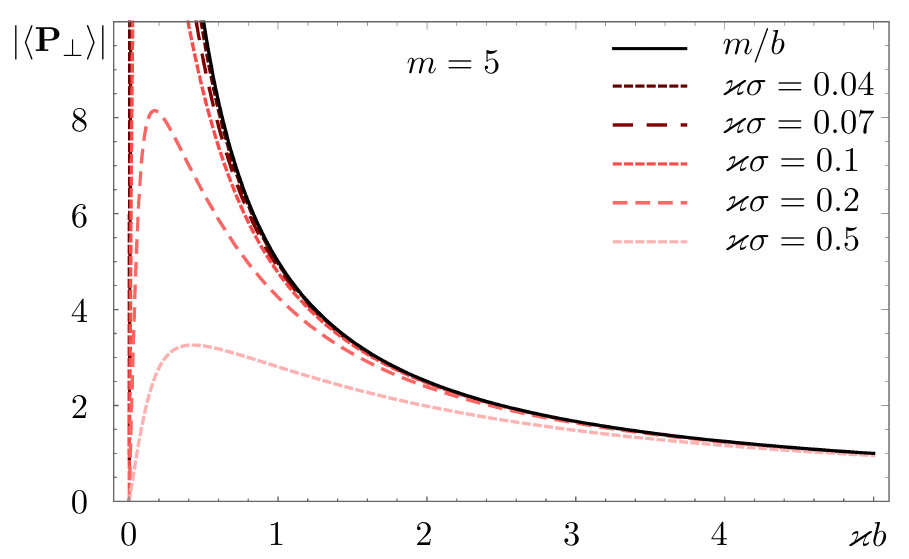}
	\caption{Transverse momentum of the final state expressed in units of $\varkappa$ as a function of $\varkappa b$
		up to $\varkappa b = 1$ (the upper plot) and up to $\varkappa b = 5$ (the lower plot) 
		for various localization parameters $\sigma$.
		The value of $m = 5$.
	In each plot, the uppermost black line shows the superkick phenomenon in the limit $\sigma \to 0$.}
	\label{fig-Pt}
\end{figure}

In order to explore how the superkick phenomenon emerges and disappears for finite $\sigma$,
we evaluated the transverse momentum $\langle \bP_\perp\rangle$ numerically using the exact
expression for the integral ${\cal J}$. The results are shown in Fig.~\ref{fig-Pt} for $m = 5$
and several values of $\sigma$, from very small, $\sigma \varkappa = 0.04$ to moderately small $\sigma \varkappa = 0.5$.
The uppermost black curve shows the expected superkick behavior $m/b = \varkappa\cdot (m/\varkappa b)$.

We see that for small $\sigma$, with $b \gg \sigma$, the curves nicely approach the black superkick regime.
However, starting from $\sigma \approx 0.5/\varkappa$ (the lowest curve in Fig.~\ref{fig-Pt}), the superkick behavior does not resemble 
the semiclassical Eq.~\eqref{mb}: the $b$ dependence is rather flat.
At small $b$, it still produces a mild superkick, as $\langle \bp_\perp \rangle$ is larger than $\varkappa$,
but the curve does not display the pronounced $1/b$ enhancement characteristic to the semiclassical expectation.

At small values of $\sigma$, when the $1/b$ dependence is clear, we can observe that the maximal value of $\langle \bP_\perp\rangle$
which is attained at $b\approx \sigma$, lies significantly below the semiclassical curve.
This factor is about 3 for the example shown, $m = 5$, and slowly grows for larger $m$, reaching about 10 for $m=50$.
One can also notice that, in order for $\langle \bP_\perp\rangle$ to be close (say, within 20\%) to the semiclassical
superkick expectation, the value of $\sigma$ must be smaller than $b/m$, meaning that the compact 
Gaussian wave packet must contain sufficiently high momenta for a given $m$ and $b$.

\section{Relation with previous works}\label{section-comments}

Our qualitative and quantitative observations are in general agreement with the findings of \cite{barnett2013superkick}.
Thus, although the approach of \cite{barnett2013superkick} is not fully quantum, we confirm that it is adequate for the problem 
of a localized atom in the vicinity of an optical vortex.
These results highlight the necessity of treating the localized particle as a compact wave packet
rather than a classical particle which, by definition, is pointlike and {\em stays} pointlike. 
Without (transverse) wave packets and their localization scale $\sigma$, 
neither can the superkick phenomenon be properly understood nor its observability limits be detected.

We showed that, in order for the superkick to be close to the semiclassical prediction, 
$\sigma$ must be smaller than $b/m$, which in turn must be smaller than $\varkappa$.
However, the smaller $\sigma$, the faster is spreading of the Gaussian wave packet and the shorter is the focal length $L_\sigma$.
This leads to a new requirement for the scattering process occurring in free space:
not only must it involve a particle localized to a fraction of transverse ring radius of the twisted beam $m/\varkappa$
and placed with high accuracy close to the beam axis, but also the interaction itself must be longitudinally localized 
to within the focal length $L_\sigma$, or alternatively, occur over a very short time.

Satisfying these conditions may be feasible in experiments with trapped atoms or ions placed in the vicinity of an optical vortex.
	Indeed, trapped atom or ion wavefunction can be localized to submicron (in fact sub-100 nm) sizes and manipulated with comparable accuracy.
For example, the recent paper \cite{Drechsler:2021} reports optical superresolution sensing of an individual trapped ${}^{40}$Ca$^+$ ion, 
whose wavefunction size was measured to be about 40~nm, in agreement with a less direct sideband spectroscopic evaluation.
At the same time, the transverse size of the first Bessel ring $1/\varkappa$, which is larger than the optical wavelength,
can reach microns.
When released from the trap, the wave packet begins to spread, but the spreading time may be large 
compared to the interaction with a short twisted light pulse.
For example, a heavy atom with $m = 100$ GeV localized in a trap to a region of $\Delta x = 10$ nm size 
will begin to spread when released and double its extent on the timescale $\tau \approx m(\Delta x)^2/\hbar \sim 100$ ns. 
Thus, nanosecond-long light pulses effectively probe a frozen wavepacket.

	Observing superkick and related effects in nuclear and particle physics processes
such as deuteron photodisintegration or $\Delta$ baryon photoproduction by twisted gamma rays 
as suggested in \cite{Afanasev:2020nur,Afanasev:2021fda} is very challenging.
For that, one would need not only to produce twisted gamma beam but also to focus it to sub-picometer 
or even femtometer focal spots, localize the target proton or deuteron wavefunction to a femtometer-scale size and 
manipulate the beam and the target with comparable precision. 
Besides, a free evolving proton wavefunction of the initial localization size of the order of femtometer 
will rapidly expand doubling its size over typical nuclear times scale of $\sim 10^{-22}$ s.
Thus, one needs either to arrange for the hard twisted photon to interact with the proton during this short time,
or to find a way to keep the proton localized.
Achieving this level of control is challenging and requires dedicated experimental efforts.

\section{Conclusions and outlook}

	In summary, we rederived within the QFT framework the superkick phenomenon predicted in \cite{barnett2013superkick}. 
Since in our calculations we treated both initial particles---the vortex state and the compact Gaussian state---as freely propagating wavepackets,
we included the full energy-momentum conservation law, which led to non-trivial effects when inserted between the initial and final states. 
On the way to the final result, we encountered several apparent paradoxes, which we resolved with qualitative insights 
and quantitative calculations. We believe that this part of our work, by itself, is a valuable result  
as it clarifies several aspects of vortex states scattering which may at first seem puzzling.

Concerning the superkick phenomenon, we stressed that its observability relies on a hierarchy of several length scales, 
see Eqs.~\eqref{scales} and \eqref{ell-choice}, and we explored the roles of these scales. 
Our QFT results confirm the interpretation and dependences of \cite{barnett2013superkick}.
In particular, they highlight the critical role of the wave packet localization parameter $\sigma$.
In the semiclassical language of pointlike particles, without taking into account the transverse localization 
and evolution of the wave packets, the superkick cannot be properly understood.

The requirements on $\sigma$ and $b$ needed for observation of the superkick and other related phenomena 
are readily met for atoms or ions interacting with optical vortex, as it was already noted in previous works on the subject.
However we remarked that observing similar effects in the nuclear and high-energy physics domains as 
recently proposed and discussed in \cite{Afanasev:2020nur,Afanasev:2021fda}
place severe requirements on target localization and alignment, which are not yet achievable with the present-day technology.

With our results and discussion, we believe that the effect of superkick is not as surprising 
as it may originally appear. Still, there remains room for improvement.
Here, we used exact monochromatic Bessel beams and, running into an artifact, devised
a somewhat artificial scheme for limiting the longitudinal range of interaction.
It would be interesting to see whether more realistic schemes 
such as Laguerre-Gaussian vs Gaussian beam collision \cite{Karlovets:2020odl} which
are more computationally heavy but free from singularities lead to the same results.
Once a more realistic treatment is constructed, one can turn to specific processes
and investigate, in particular, their energy behavior and the energy threshold shift 
proposed in \cite{Afanasev:2020nur,Afanasev:2021fda}. 

\section*{Acknowledgments}

We thank Pengcheng Zhao and Liping Zou for many stimulating discussions and Dmitry Karlovets for valuable comments.
This work was supported by the National Natural Science Foundation of China 
(Grant No. 11975320) and the Fundamental Research Funds for the Central Universities, Sun Yat-sen University.

\appendix

\section{The two-particle phase space}\label{appendix-2particles}

The expression \eqref{sigma-general} for a generic $2\to 2$ scattering cross section for monochromatic but nonplane-wave initial states 
involves the two-particle phase space without the three-momentum constraints:
\begin{equation}
d\Gamma' = \delta(E_0 - E_1' - E_2') \, \frac{d^3 \bk'_1}{2E'_1} \frac{d^3 \bk'_2}{2E'_2}\,,\label{dGam}
\end{equation}
where $E_0 = E_1 + E_2$ is fixed. The energy delta function can be used to integrate one of the variables
leading to a five-dimensional phase space distribution.
In this section, we find a convenient way to perform it.\footnote{Technically, 
	this expression is similar to the three-particle phase space in a usual plane-wave collision,
but we will not rely on this analogy.}

In plane-wave scattering processes, when analyzing final particle distributions, 
one is free to select a convenient reference frame where the analysis simplifies.
In our case, however, we cannot perform any boost since it will spoil the monochromaticity of the colliding beams.
We are constrained to work in a generic frame; the only simplification is related to the natural choice of axis $z$
and the transverse momenta orthogonal to it.
Switching from $d^3 \bk'_1 d^3 \bk'_2$ to $d^2 \bk'_{1\perp} dk'_{1z}\, d^3 \bP$,
where $\bP = \bk'_1 + \bk'_2$, 
we perform the $dk'_{1z}$ integration to remove the delta-function while keeping 
$\bk'_{1\perp}$, $\bP_\perp$, and $P_z$ as independent variables.

If $m_1$ and $m_2$ are the masses of the two final particles, we can define their ``transverse masses'' as
\begin{equation}
m_{1\perp}^2 = m_1^2 + \bk_{1\perp}^{\prime 2} = E_1^{\prime 2} - k_{1z}^{\prime 2}\,, \quad 
m_{2\perp}^2 = m_2^2 + (\bP_\perp-\bk'_{1\perp})^2 = E_2^{\prime 2} - (P_z-k'_{1z})^{2}\,.
\end{equation}
In a similar way, we define the invariant mass of the final two-particle system defined as $M_{inv}^2 = E_0^2 - \bP^2$
and denote its ``transverse mass''
\begin{equation}
M_\perp^2 = M_{inv}^2 + \bP_\perp^2 = E_0^2 - P_z^2\,.
\end{equation} 
The energy delta-function can be now expressed as a function of $k'_{1z}$:
$$
\delta(E_0 - E_1' - E_2') = \delta(f(k'_{1z}))\,, \quad 
f(k'_{1z}) = E_0 - \sqrt{m_{1\perp}^2 + k_{1z}^{\prime 2}} - \sqrt{m_{2\perp}^2 + (P_z-k'_{1z})^2} 
$$
and integrated out as
\begin{equation}
\int dk'_{1z} \frac{\delta(E_0 - E_1' - E_2')}{E_1'E_2'} = \int dk'_{1z} \frac{\delta(f(k'_{1z}))}{E_1'E_2'} 
= \sum_{k'_{1z} = k^*_{1z}} \frac{1}{|df/dk'_{1z}|E_1'E_2'}\,,\label{delta-trans}
\end{equation}
where $k^*_{1z}$ are the solutions of the equation $f(k'_{1z}) = 0$. The solutions of this equations
can be expressed as
\begin{eqnarray}
k^*_{1z} &=& \frac{1}{2}P_z\left(1+\frac{m_{1\perp}^2 - m_{2\perp}^2}{M_\perp^2}\right)
\pm \frac{E_0}{2M_\perp^2}\sqrt{\lambda(M_\perp^2,m_{1\perp}^2,m_{2\perp}^2)}\,,\nonumber\\
P_z - k^*_{1z} &=& \frac{1}{2}P_z\left(1-\frac{m_{1\perp}^2 - m_{2\perp}^2}{M_\perp^2}\right)
\mp \frac{E_0}{2M_\perp^2}\sqrt{\lambda(M_\perp^2,m_{1\perp}^2,m_{2\perp}^2)}\,,
\end{eqnarray}
where
\begin{equation}
\lambda(M_\perp^2,m_{1\perp}^2,m_{2\perp}^2) = M_\perp^4 + m_{1\perp}^4 + m_{2\perp}^4 
- 2M_\perp^2 m_{1\perp}^2 - 2M_\perp^2 m_{2\perp}^4 - 2m_{1\perp}^2m_{2\perp}^2\,.\label{lambda}
\end{equation}
This solution exists only if the transverse masses satisfy the following inequality:
\begin{equation}
M_\perp \ge m_{1\perp} + m_{2\perp}\,,
\end{equation}
which places a constraint on the possible values of $\bk'_{1\perp}$, $\bP_\perp$, and $P_z$.
The value of $k'_{1z} = k^*_{1z}$ corresponds to the following energies of the final particles:
\begin{eqnarray}
E'_{1} &=& \frac{1}{2}E_0\left(1+\frac{m_{1\perp}^2 - m_{2\perp}^2}{M_\perp^2}\right)
\pm \frac{P_z}{2M_\perp^2}\sqrt{\lambda(M_\perp^2,m_{1\perp}^2,m_{2\perp}^2)}\,,\nonumber\\
E'_{2} &=& \frac{1}{2}E_0\left(1-\frac{m_{1\perp}^2 - m_{2\perp}^2}{M_\perp^2}\right)
\mp \frac{P_z}{2M_\perp^2}\sqrt{\lambda(M_\perp^2,m_{1\perp}^2,m_{2\perp}^2)}\,,\nonumber\\
\end{eqnarray}
The derivative $|df/dk'_{1z}|$ entering \eqref{delta-trans} can be expressed as $|k^*_{1z}/E'_1 - (P_z - k^*_{1z})/E'_2|$
so that
\begin{equation}
\left|\frac{df}{dk'_{1z}}\right|E'_1E'_2 = |k^*_{1z} E_0 - P_zE'_1| 
= \frac{1}{2}\sqrt{\lambda(M_\perp^2,m_{1\perp}^2,m_{2\perp}^2)}\,.
\end{equation}
This result allows us to finally represent the final particle phase space as
\begin{equation}
d\Gamma' = \sum_{k^*_{1z}} \frac{d^2 \bk'_{1\perp} \, d^2 \bP_\perp \, dP_z}{2\sqrt{\lambda(M_\perp^2,m_{1\perp}^2,m_{2\perp}^2)}}\,.\label{dGam2}
\end{equation}
Here, we keep the summation over the two solutions for $k^*_{1z}$ because the values of the invariant amplitude ${\cal M}$,
and consequently ${\cal I}$, can differ for the two final state momenta.

Let  us understand the behavior of the $1/\sqrt{\lambda}$ factor within the available phase space.
In our kinematic situation, we have in mind a high-energy collision with large $M_{inv}^2$ and small $\bP_\perp^2$.
Thus, when studying $\bP_\perp$ distribution, we can safely assume that $M_\perp^2$ stays almost constant and approximately equal to
$M_{inv}^2 \approx E_0^2-P_{z0}^2$, where $P_{z0}= p_{1z}+p_{2z}$. The values of $m_{1\perp}^2$ and $m_{2\perp}^2$ depend on the final momenta,
which are large, $|\bk'_{1\perp}|\gg |\bP_\perp|$. In particular, in the vast majority of the phase space we have
$m_{2\perp}^2 = m_2^2 + (\bP_\perp-\bk'_{1\perp})^2 \approx m_2^2 + \bk_{1\perp}^{\prime 2}$, 
so that $\lambda$ shows almost no dependence on $\bP_\perp$ or $P_z$
but is a function of $\bk'_{1\perp}$ only. We conclude that the final phase space can be approximately factorized:
\begin{equation}
d\Gamma' = \mbox{relative motion} \times d^2 \bP_\perp \, dP_z\,.\label{dGam3}
\end{equation}
This factorized form will allows us to evaluate the total transverse momentum of the final state
for a generic value of $\bk'_{1\perp}$.

\section{Absence of the superkick due to non-decoupling of the Bessel and Gaussian states}\label{appendix-absence}

Here, we continue calculating the integral \eqref{main-2} for the exact Bessel and Gaussian state scattering:
\begin{eqnarray}
{\cal I} &=& \int d^3\bk_1 d^3\bk_2 \, \delta(k_{1z} - p_{1z})\delta(|\bk_{1\perp}| - \varkappa) e^{i m \varphi_1}\cdot 
\, e^{-\frac{(\bk_2 - \bp_2)^2\sigma^2}{2}} \delta(\bk_2^2 + M^2 - E_2^2)\, e^{-i\bk_{2\perp}\bb_\perp} 
\delta^{(3)}(\bk_1 + \bk_2 - \bP) \cdot {\cal M}\,.\nonumber 
\end{eqnarray}
Five out of six delta functions can be taken care of by the integrations leading to 
\begin{eqnarray}
{\cal I} &=& \varkappa \int d\varphi_1 
\exp\left[i m \varphi_1 - i (\bP_\perp-\bk_{1\perp})\bb_\perp - \frac{(\bP_\perp-\bk_{1\perp})^2\sigma^2}{2}
-\frac{(P_{z} - p_{1z}-p_{2z})^2\sigma^2}{2}\right] \nonumber\\
&&\qquad \times\ 
\delta\left[(\bP_\perp-\bk_{1\perp})^2 + (P_{z} - p_{1z})^2 - p_{2z}^2\right]\cdot {\cal M}\,.\label{main-4}
\end{eqnarray}
Here, the only remaining variable is the azimuthal angle of $\bk_{1\perp}$, 
with $|\bk_{1\perp}|=\varkappa$ already fixed.
Also, in the last delta-function we replaced $E_2^2-M^2$ with $p_{2z}^2$.
To perform the remaining integration, we denote the azimuthal angles of $\bP_\perp$ and $\bb$ as
$\varphi_P$ and $\varphi_b$, respectively, and their difference as $\varphi_{bP} = \varphi_b - \varphi_P$.
Then we introduce the shifted variable $\tilde \varphi_1 = \varphi_1 - \varphi_P$ and rewrite \eqref{main-4} as
\begin{equation}
{\cal I} = \varkappa\, e^{im\varphi_P} e^{ - i\bP_\perp\bb_\perp} e^{- \sigma^2 |p_{2z}|(P_z - p_{1z}-p_{2z})} 
\int d\tilde\varphi_1 e^{i m \tilde\varphi_1 + i \varkappa b \cos(\tilde\varphi_1-\varphi_{bP})} 
\delta(a - c\cos\tilde\varphi_1)\cdot {\cal M}\,,\label{main-5}
\end{equation}
where
\begin{equation}
a = \bP_\perp^2 + \varkappa^2 + (P_z-p_{1z})^2 - p_{2z}^2\,, \quad c = 2 |\bP_\perp|\varkappa\,.\label{a-c}
\end{equation}
The integral is nonzero only when $|a| < c$, which places constraints on possible values of $\bP_\perp$ and $P_z$.
The delta function fixes the angle $\tilde\varphi_1 = \pm \varphi_*$, where $\cos\varphi_* = a/c$.
The value of the invariant amplitude ${\cal M}$ can, in principle, be different at these two values of $\tilde\varphi_1$,
which may lead to interesting effects similar to those predicted in double-twisted electron scattering \cite{Ivanov:2016oue}.
However since we are interested here in kinematic effects, we assume that ${\cal M}$ takes the same value 
at these two angles $\tilde\varphi_1$. This allows us to simplify the integral as
\begin{equation}
{\cal I} = e^{im\varphi_P} e^{ - ib\cos\varphi_{bP} (|\bP_\perp| - a/2|\bP_\perp|)}
e^{- \sigma^2 |p_{2z}|(P_z - p_{1z}-p_{2z})} 
\cdot \frac{\cal M}{|\bP_\perp|\sin\varphi_*}\cos(m\varphi_* + \varkappa b \sin\varphi_* \sin \varphi_{bP})\,.\label{main-6}
\end{equation}
This expression represents the outgoing wavefunction of the center of mass motion of the final state $\Psi_f(\bP)$
in momentum representation. Its modulus squared 
\begin{equation}
|{\cal I}|^2 = e^{- 2\sigma^2 |p_{2z}|\Delta P_z } 
\cdot \frac{|{\cal M}|^2}{\bP_\perp^2\sin^2\varphi_*}\cos^2(m\varphi_* + \varkappa b \sin\varphi_* \sin \varphi_{bP})\label{main-7}
\end{equation}
defines the differential cross section \eqref{sigma-general}.

\begin{figure}[!h]
	\centering
	\includegraphics[width=0.4\textwidth]{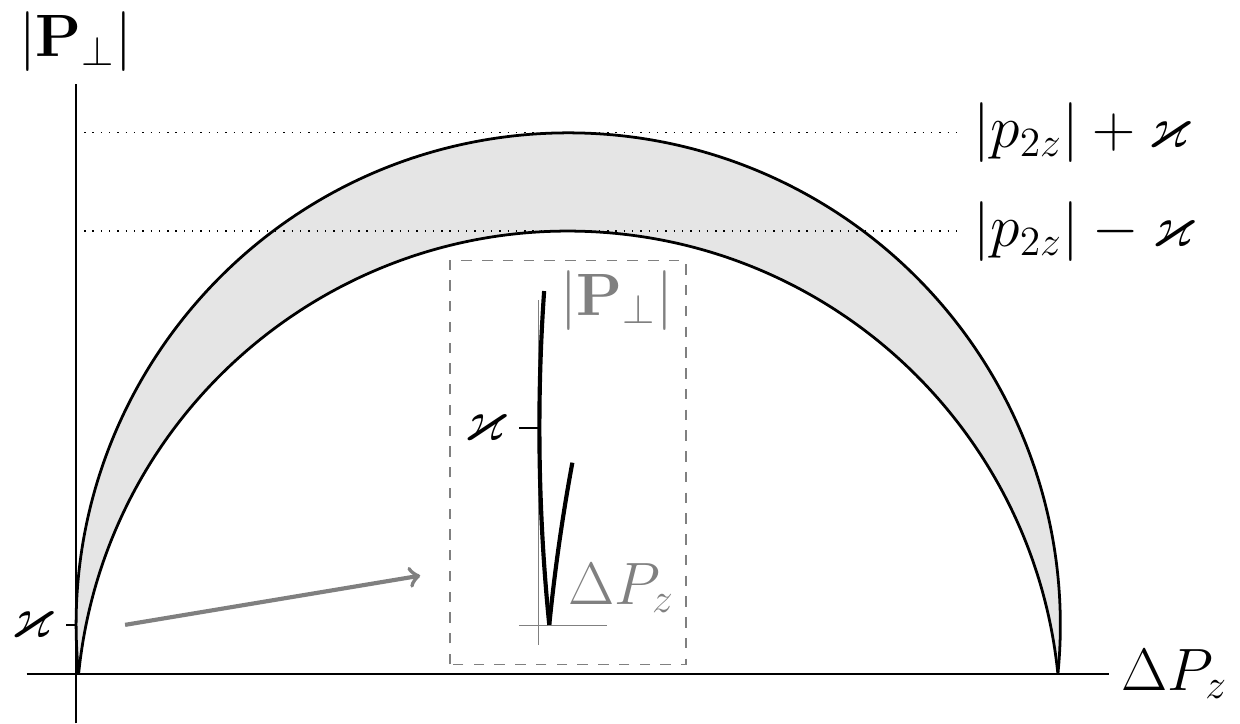}
	\caption{The crescent-like region on the $(|\bP_\perp|,\Delta P_z)$ plane 
		defined by the kinematical constraints in Eq.~\eqref{bP-Pz-plane-exact}. 
		The inset shows the zoomed region near the origin. The plot corresponds to $\varkappa = 0.1|p_{2z}|$.
		}
	\label{fig-arcs}
\end{figure}

	Let us understand the kinematic constraints on $P_z$ and $\bP_\perp$
which arise from the restriction $|a| < c$ in \eqref{a-c}.
To this end, we use $\Delta P_z = P_z - p_{1z} +|p_{2z}|$ defined in Eq.~\eqref{DeltaPz}, 
which shows how the value of $P_z$ differs from the paraxial limit value, and rewrite 
\begin{equation}
(P_z-p_{1z})^2 - p_{2z}^2 = \Delta P_z (2p_{2z}+\Delta P_z)\,. 
\end{equation}
The two conditions $a<c$ and $-a< c$ take the following form:
\begin{equation}
(|\bP_\perp|-\varkappa)^2 < 2 |p_{2z}|\Delta P_z - (\Delta P_z)^2 < (|\bP_\perp|+\varkappa)^2\,,\label{bP-Pz-plane-exact} 
\end{equation}
from which one immediately sees that $\Delta P_z \ge 0$.
These two conditions cut out a crescent-like region on the $(\Delta P_z, |\bP_\perp|)$ plane bounded by two arcs as shown in Fig.~\ref{fig-arcs}
(this plot corresponds to $\varkappa = 0.1|p_{2z}|$).
It is inside this region that the function $|{\cal I}|^2$ of \eqref{main-7} is defined.
The maximal value of $\Delta P_z$ is $2|p_{2z}|$, the maximal value of $|\bP_\perp|$ is $|p_{2z}|+\varkappa$.
In the region of small $\Delta P_z$ and $|\bP_\perp|$, the arcs do not reach the origin but meet at the $|\bP_\perp|=0$ axis
at the value $\Delta P_z \approx \varkappa^2/2|p_{2z}|$, shown in the inset of Fig.~\ref{fig-arcs}.
The larger arc touches the $\Delta P_z=0$ axis at $|\bP_\perp|=\varkappa$.

The cross section is not distributed homogeneously over this crescent-like region.
The first exponential factor in $|{\cal I}|^2$ in Eq.~\eqref{main-7} together ith the strong inequality $\sigma |p_{2z}| \gg 1$ 
make it clear that most of the cross section comes from the origin of small $\Delta P_z$. 
In this region, we can neglect $(\Delta P_z)^2$ when compared to $2 |p_{2z}|\Delta P_z$ and replace
the exact inequalities of Eq.~\eqref{bP-Pz-plane-exact} with the approximation Eq.~\eqref{bP-Pz-plane}
given in the main text. The typical ranges of $\Delta P_z$ and $|\bP_\perp|$ are expected to be
\begin{equation}
\Delta P_z \lsim \frac{1}{\sigma^2 |p_{2z}|}\,, \quad
|\bP_\perp| \lsim \frac{1}{\sigma}\,,\quad
\Delta P_z \ll |\bP_\perp|\,.\label{DeltaPz-Pperpb}
\end{equation} 
The last relation confirms the expectation that the total momentum of the final state is predominantly transverse,
and that, for a well-localized Gaussian beam with $\sigma \ll 1/\varkappa$, 
it is mostly driven by the transverse momenta of the initial compact Gaussian state, 
not by the transverse momentum of the Bessel twisted state.

Inside this small-$\Delta P_z$, small-$|\bP_\perp|$ region, the cross section 
exhibits strong variations.
As we pick up a $|\bP_\perp|$ value and scan the allowed region horizontally by varying $\Delta P_z$,
the value of $\varphi_*$ varies from 0 to $\pi$.
So, at the boundaries, we expect to observe endpoint divergences $\propto 1/\sin^2\varphi_*$, 
while inside this region we expect to see the oscillations of the cosine squared.

Next, we need to compute the average transverse momentum of the final state via \eqref{average-P-def}.
Let us focus on the integration over $\Delta P_z$ for any given $\bP_\perp$. Since
$$
d \Delta P_z \approx \sin\varphi_*d\varphi_*\, \frac{|\bP_\perp|\varkappa}{|p_{2z}|}\,,
$$
we see that the integration over $\Delta P_z$ diverges at the boundaries:
\begin{equation}
\int_0^\pi \frac{d\varphi_*}{\sin\varphi_*}\cos^2(m\varphi_* + \varkappa b \sin\varphi_* \sin \varphi_{bP})\,.
\end{equation}
is undefined. This divergence is omnipresent in all scattering processes with exact Bessel twisted states
\cite{Jentschura:2010ap,Jentschura:2011ih,Ivanov:2011kk,Karlovets:2012eu} and is driven by the fact that
such states are not normalizable.
To make the integrals finite, one must define a regularization scheme. 
One possibility is to normalize a Bessel twisted state in a large but finite cylindrical volume
of length $L$ and radius $R$ \cite{Jentschura:2010ap}. Another option is to assume that 
the twisted state is not an exact Bessel beam with fixes $\varkappa$ but is represented 
by some distribution over a small range of $\varkappa$ \cite{Ivanov:2011bv}.
Whatever prescription for regularizing the divergent $\int d\varphi_*/\sin\varphi_*$ integration, 
we obtain a finite result. Even if it is enhanced by a large logarithm, 
it will be the same in the numerator and denominator of \eqref{average-P-def} and the potentially large factors cancel.

We conclude from this discussion that, via a suitable regularization procedure, it is possible to make the $P_z$
integrals finite, with the largest contributions coming from the regions near the boundaries,
that is, where $\varphi_*$ is close to 0 or $\pi$.

Next, we switch to the transverse momentum integration $d^2\bP_\perp = d\bP_\perp^2 d\varphi_P/2$. 
We remind the reader that our goal is to verify the phenomenon of superkick.
The key expression is
\begin{equation}
\int d\varphi_P \bP_\perp \cos^2(m\varphi_* + \varkappa b \sin\varphi_* \sin \varphi_{bP})
\end{equation}
which we need to analyze for $\varkappa b \ll 1$. Let us select axis $x$ along $\bb_\perp$ as in Fig.~\ref{fig-semiclassical}.
Then $\varphi_{bP} = \varphi_P$ and $\bP_\perp = |\bP_\perp|(\cos \varphi_P,\, \sin \varphi_P)$.
We see that only $P_y$ gives a nonvanishing result, which, at $\varkappa b \ll 1$, can be written as
\begin{equation}
\frac{1}{2}|\bP_\perp|\int d\varphi_P \sin\varphi_P \left[-\sin(2m\varphi_*)\cdot 2\varkappa b \sin\varphi_* \sin\varphi_P\right]
= - \pi |\bP_\perp| \varkappa b \cdot \sin(2m\varphi_*) \sin\varphi_*\,.\label{kick-2}
\end{equation} 
This result {\em does not} confirm the superkick phenomenon.
We see that the transverse momentum transfer is proportional to $b$, not $1/b$, as expected from the semiclassical analysis of
the pointlike probe recoil.
Moreover, its value is additionally suppressed by the two sine factors computed for $\varphi_*$ near $0$ or $\pi$.

Thus, we seem to run into yet another paradox: the exact QFT treatment 
of the Gaussian vs. Bessel beam scattering fails to reproduce the superkick phenomenon
obtained with semiclassical considerations. 

In the main text, we show that this conclusion is an artifact of the inappropriate setting 
we just used and demonstrate how the problem can be cured.


\begin{thebibliography}{99}

\bibitem{rubinsztein2016roadmap}
H.~Rubinsztein-Dunlop, A.~Forbes, M.~V. Berry, M.~R. Dennis, D.~L. Andrews,
M.~Mansuripur, C.~Denz, C.~Alpmann, P.~Banzer, T.~Bauer, et~al.,
Journal of Optics \textbf{19}, 013001 (2017).

\bibitem{babiker2019atoms}
M.~Babiker, D.~L. Andrews, V.~E. Lembessis,
Journal of Optics \textbf{21}, 013001 (2019).

\bibitem{forbes2021structured}
A.~Forbes, M.~de~Oliveira, M.~R. Dennis,
Nature Photonics \textbf{15}, 4, 253 (2021).

\bibitem{Allen:1992zz}
L.~Allen, M.~W. Beijersbergen, R.~J.~C. Spreeuw, J.~P. Woerdman,
Phys. Rev. A \textbf{45}, 8185 (1992).

\bibitem{Paggett:2017}
M.~J. Padgett,
Opt. Express \textbf{25}, 11265 (2017).

\bibitem{Knyazev:2019}
B.~A. Knyazev, V.~G. Serbo, 
Phys. Uspekhi \textbf{61}, 449 (2018).

\bibitem{andrews2012angular}
D.~L. Andrews, M.~Babiker,
\emph{The angular momentum of light},
Cambridge University Press (2012).

\bibitem{babiker1994}
M.~Babiker, W.~L.~Power, and L.~Allen
Phys. Rev. Lett. \textbf{73}, 1239 (1994).
	
\bibitem{babiker2002orbital}
M.~Babiker, C.~Bennett, D.~Andrews, L.~D. Romero,
Phys. Rev. Lett. \textbf{89}, 14, 143601 (2002).

\bibitem{Afanasev:2014}
A. Afanasev, C.~E. Carlson, and A. Mukherjee, 
J. Opt. Soc. Am. B \textbf{31}, 11, 2721 (2014).

\bibitem{schmiegelow2016transfer}
C.~T. Schmiegelow, J.~Schulz, H.~Kaufmann, T.~Ruster, U.~G. Poschinger,
F.~Schmidt-Kaler,
Nature communications \textbf{7}, 12998 (2016).

\bibitem{afanasev2018experimental}
A.~Afanasev, C.~E. Carlson, C.~T. Schmiegelow, J.~Schulz, F.~Schmidt-Kaler,
M.~Solyanik,
New Journal of Physics \textbf{20}, 2, 023032 (2018).

\bibitem{solyanik2019excitation}
M.~Solyanik-Gorgone, A.~Afanasev, C.~E. Carlson, C.~T. Schmiegelow,
F.~Schmidt-Kaler,
J. Opt. Soc. Am. B \textbf{36}, 3, 565 (2019).

\bibitem{schulz2019modification}
S.-L. Schulz, S.~Fritzsche, R.~A. M{\"u}ller, A.~Surzhykov,
Phys. Rev. A \textbf{100}, 4, 043416 (2019).

\bibitem{barnett2013superkick}
S.~M. Barnett, M.~Berry,
Journal of Optics \textbf{15}, 12, 125701 (2013).

\bibitem{Afanasev:2020nur}
A.~Afanasev, C.~E.~Carlson and A.~Mukherjee,
Phys. Rev. Res. \textbf{3}, no.2, 023097 (2021).

\bibitem{Berry2013fivemomenta}
M.~V. Berry, 
Eur. J. Phys. \textbf{34}, 1337 (2013).

\bibitem{Afanasev:2021fda}
A.~Afanasev and C.~E.~Carlson,
Ann. Phys. (Leipzig) \textbf{2021}, 2100228.


\bibitem{Peskin}
M.~E.~Peskin and D.~V.~Schroeder,
``An Introduction to quantum field theory,''
CRC Press (1995).


\bibitem{Kotkin:1992bj}
G.~L.~Kotkin, V.~G.~Serbo and A.~Schiller,
Int. J. Mod. Phys. A \textbf{07}, 4707-4745 (1992).

\bibitem{Karlovets:2016jrd}
D.~Karlovets,
JHEP \textbf{2017}, 49 (2017).

\bibitem{Karlovets:2020odl}
D.~V.~Karlovets and V.~G.~Serbo,
Phys. Rev. D \textbf{101}, no.7, 076009 (2020).

\bibitem{Jentschura:2010ap}
U.~D.~Jentschura and V.~G.~Serbo,
Phys. Rev. Lett. \textbf{106}, no.1, 013001 (2011).

\bibitem{Jentschura:2011ih}
U.~D.~Jentschura and V.~G.~Serbo,
Eur. Phys. J. C \textbf{71}, 1571 (2011).

\bibitem{Karlovets:2012eu}
D.~V.~Karlovets,
Phys. Rev. A \textbf{86}, 062102 (2012).

\bibitem{Ivanov:2011kk}
I.~P.~Ivanov,
Phys. Rev. D \textbf{83}, 093001 (2011).

\bibitem{Ivanov:2016oue}
I.~P.~Ivanov, D.~Seipt, A.~Surzhykov and S.~Fritzsche,
Phys. Rev. D \textbf{94}, 076001 (2016).

\bibitem{Ivanov:2011bv}
I.~P.~Ivanov and V.~G.~Serbo,
Phys. Rev. A \textbf{84}, 033804 (2011).

\bibitem{Karlovets:2021gcm}
D.~V.~Karlovets, V.~G.~Serbo and A.~Surzhykov,
Phys. Rev. A \textbf{104}, 023101 (2021).

\bibitem{integrals}
D.~Zwillinger, V.~Moll, I.~S.~Gradshteyn and I.~M.~Ryzhik,
Table of Integrals, Series, and Products (Eighth Edition),
Academic Press (2014).

\bibitem{Drechsler:2021}
M.~Drechsler, S.~Wolf, C.~T.~Schmiegelow, and F.~Schmidt-Kaler,
Phys. Rev. Lett. \textbf{127}, 143602 (2021).

\end{thebibliography}
\end{document}